\documentclass[11pt]{article}

\usepackage{amsmath}

\usepackage[preprint]{acl}

\usepackage{times}
\usepackage{latexsym}

\usepackage[T1]{fontenc}

\usepackage[utf8]{inputenc}

\usepackage{microtype}

\usepackage{inconsolata}

\usepackage{graphicx}

\usepackage{listings}

\usepackage{caption}
\DeclareCaptionLabelFormat{nolabel}{}
\lstdefinelanguage{plain}{
  morestring=[b]{},
}

\title{QRAFTI: An Agentic Framework for \\ Empirical Research in Quantitative Finance}

\author{
  \textbf{Terence Lim\textsuperscript{1, 2}},
  \textbf{Kumar Muthuraman\textsuperscript{2}},
  \textbf{Michael Sury\textsuperscript{2}}
\\
  \textsuperscript{1}Graphen Inc.,
  \textsuperscript{2}The University of Texas at Austin
}

\begin{document}
\maketitle
\begin{abstract}
Quantitative financial research is a multi-stage process that typically involves generating ideas grounded in economic intuition, acquiring and cleaning data, constructing signals and factors, backtesting portfolios, adjusting for risk, evaluating robustness to model uncertainty and data-mining bias, and producing research reports that enable peer verification.
Recent advances in agentic AI---that is, LLM-centered systems that can plan, invoke external tools, and revise actions in response to observed outputs---create an opportunity to assist parts of this workflow while potentially improving reproducibility and explainability.
We introduce QRAFTI, a multi-agent framework intended to emulate parts of a quantitative research team and support equity factor research on large financial panel datasets. It integrates an empirical research toolkit for panel data with MCP servers that expose data access, factor construction, and custom coding operations as callable tools. Through this architecture, QRAFTI can help replicate established factors, formulate and test new signals, and generate standardized research reports accompanied by narrative analysis and computational traces.
Our initial evaluations suggest that, on multi-step empirical research tasks, the system offers better performance and explainability using chained tool calls and reflection-based planning than dynamic code generation alone.
\end{abstract}

\let\oldsection\section
\let\oldsubsection\subsection
\let\chapter\oldsection
\let\section\oldsubsection
\let\subsection\subsubsection

\chapter[Introduction]{Introduction}

Systematic equity investing involves discovering and validating predictors of the cross-section of stock returns---commonly referred to as \emph{characteristics}, \emph{signals}, or \emph{factors}. A contemporary empirical finance workflow commonly includes:
(1) generating signals and factor portfolios grounded in economic intuition,
(2) evaluating them on large \emph{panel} datasets spanning broad cross-sectional universes over multiple decades, and
(3) conducting robustness and implementability checks.
This research process is data-intensive and often sensitive to implementation choices: seemingly minor decisions, such as universe screens, lag conventions, breakpoint definitions, and weighting rules, can materially affect empirical results. The resulting ``factor zoo'' \citep{cochrane2011presidential} reflects both a large hypothesis space and a substantial body of published predictors, together with uneven standards for replication and implementability \citep{ChenZimmermann2021, JensenKellyPedersen2023, novy-marx2023assaying}.

At the same time, large language models have evolved from static text generators into tool-using, agentic systems capable of decomposing tasks, invoking external programs, and refining plans based on observed outputs \citep{yao2022react, schick2023toolformer, yang2023autogpt}. In finance, this development has motivated growing interest in domain-specific models and agent benchmarks \citep{li2024investorbench, lee2024survey, wu2023bloomberggpt, robinson2025alphagpt}.

This project brings these developments together by evaluating a
multi-agent system intended to support core components of empirical
asset-pricing research workflows, with an emphasis on reproducibility and
explainability beyond factor discovery alone.

Our design is motivated by the observation that empirical asset-pricing research is a multi-step process in which LLMs can lose track of prior results or evaluation goals during long, interdependent workflows \citep{liu-etal-2024-lost, laban2025}. In response, we adopt a multi-agent architecture with specialized roles, a constrained tool interface over panel-data operations that standardizes recurring financial computations, reflection-based planning, and structured logging and reporting that make intermediate decisions easier to inspect and reproduce. We then examine whether these design choices improve the execution of representative factor-research workflows relative to less structured configurations, while also supporting reproducibility and evaluation through computational traces and diagnostics.

We introduce QRAFTI (``Quantitative Research Assistants with Financial Tools and Intelligence''), a tool-augmented, agent-based framework\footnote{Code, documentation and an Internet Appendix supporting this work in progress are available in \url{https://github.com/terence-lim/quant-agents}} designed to reproduce core elements of an empirical research pipeline.
QRAFTI translates open-ended natural-language research requests into explicit multi-step computations by coordinating specialized agents that invoke factor-research tools and write and execute custom code when the available tools are insufficient.
It produces structured logs of relationships between intermediate and final artifacts, which can be visualized as a computation graph, and standardized research diagnostics, including controls for known factors and stratified robustness checks: these features are intended to make multi-step research pipelines easier to inspect, reproduce, and audit.

\chapter[Related Work]{Related Work}

\section{Financial asset pricing}
Asset-pricing research typically begins with economic intuition---for example, risk-based explanations, behavioral biases, or market frictions---and proposes return predictors to be tested in the cross-section \citep{Ang2014asset}. The Capital Asset Pricing Model (CAPM) links expected returns to market beta, but its empirical limitations motivated the development of multi-factor models, including the canonical market, size, and value factors \citep{fama1993common}. The Fama--French five-factor model extends this framework by incorporating profitability and investment factors, improving explanatory power in many settings \citep{fama2015five}. In practice, such models serve as
benchmarks for risk adjustment and alpha estimation,
factor-spanning tests, and
baselines for portfolio evaluation.

Two foundational empirical methodologies are:
(1) portfolio sorts \citep{fama1993common}, and
(2) cross-sectional regressions \citep{fama1973macbeth}.
Both approaches begin by analyzing the data cross-sectionally one period at a time and then studying the resulting time series of statistics, thereby mitigating the effect of cross-sectional dependence on statistical inference.

\section{Replication and the factor zoo}
The resulting empirical literature contains hundreds of proposed predictors, many of which may be fragile under replication, multiple-testing adjustments, or transaction-cost-aware evaluation \citep{harvey2016crosssection, hou2020replicating, demiguel2024multifactor, detzel2023model}. A central challenge in modern finance is distinguishing genuine risk factors from statistical artifacts generated by extensive data mining \citep{Mishra2024ReEmergence}. This difficulty is often compounded by ``$p$-hacking,'' whereby pressure to publish significant results encourages ex-post theoretical rationalizations for random patterns \citep{novy-marx2025ai}, rather than ex-ante, theory-driven hypotheses. To address this problem, researchers have argued that the conventional threshold of $|t|\approx 2$ is too permissive in large-scale exploratory settings. For example, \citet{harvey2016crosssection} propose substantially higher hurdles for validating new discoveries.

Replication-oriented datasets and open-source implementations have therefore become increasingly important in empirical asset pricing. \citet{ChenZimmermann2021} provide code and data for reproducing a large set of published predictors, while \citet{hou2020replicating} shows that many anomalies weaken under more careful implementation standards. \citet{novy-marx2023assaying} propose a standardized diagnostic framework for assessing whether a newly proposed signal is truly distinct or merely a repackaging of an existing anomaly, and whether it remains economically meaningful once practical constraints are considered. Their framework includes signal diagnostics for distributional properties and turnover, predictability checks for both raw and risk-adjusted returns, and global controls that evaluate marginal predictive power relative to a broad factor universe.

\section{LLMs and agentic systems}

Agentic AI frameworks increasingly emphasize tool integration, role specialization, and task delegation \citep{Sapkota2025}. Tool-augmented LLMs replace ``best-effort'' natural-language reasoning with grounded computation, where models can learn when and how to invoke tools in a self-supervised manner \citep{schick2023toolformer}. Multi-agent orchestration frameworks such as AutoGen and MetaGPT provide reusable interaction patterns and role definitions for complex workflows \citep{wu2023autogen, hong2023metagpt}. Other frameworks, including Reflexion and Self-Refine, rely on self-critique to help agents detect errors and improve their strategies during task execution \citep{shinn2023reflexion, madaan2023selfrefine}. Agentic systems are also being applied to scientific discovery. For example, \citet{gao2024empowering} survey AI agents in biomedical discovery, and \citet{swanson2025virtuallab} demonstrate \textit{Virtual Lab} in which an LLM ``principal investigator'' coordinates specialist agents and human collaborators to design and validate SARS-CoV-2 nanobodies.

\section{Finance-focused LLM agents}
Recent work on finance-focused LLM agents has begun to formalize how agentic systems can support investment research and trading beyond single-turn idea generation. \citet{yuan2024alphagpt2} describe a multi-agent architecture spanning alpha mining, modeling, and analysis, while \citet{cheng2024aapm} combine memory-augmented LLM analysis of financial news with traditional quantitative factors. In factor discovery, \citet{li2024fama} emphasize diversity-aware prompting and experience reuse, \citet{duan2025factormad} study iterative multi-agent critique for factor refinement, and \citet{han2026quantaalpha} propose trajectory-level evolutionary search with consistency constraints. \citet{Deng2025} studies a tool-orchestrated agent that actively gathers heterogeneous financial information and is optimized with reinforcement learning, while \citet{lopezlira2025} develops a market simulation in which prompt-defined LLM traders interact through a realistic trading environment and exhibit strategy-consistent behavior. \citet{huang2026} demonstrate a closed-loop agentic framework to iteratively discover economically interpretable factors. Benchmarking efforts such as InvestorBench formalize financial decision problems in interactive environments and evaluate multiple backbone models \citep{li2024investorbench}. While multi-agent frameworks can materially improve performance on portfolio-management analytics \citep{kundu2025multiaagentqf}, they may best be viewed as supervised research assistants rather than fully autonomous financial decision-makers \citep{Batra2025ReviewLLM}.

\chapter[Methodology]{Methodology}

\section{A multi-agent research team}
\begin{figure*}[htbp]
  \centering
  \includegraphics[width=\textwidth]{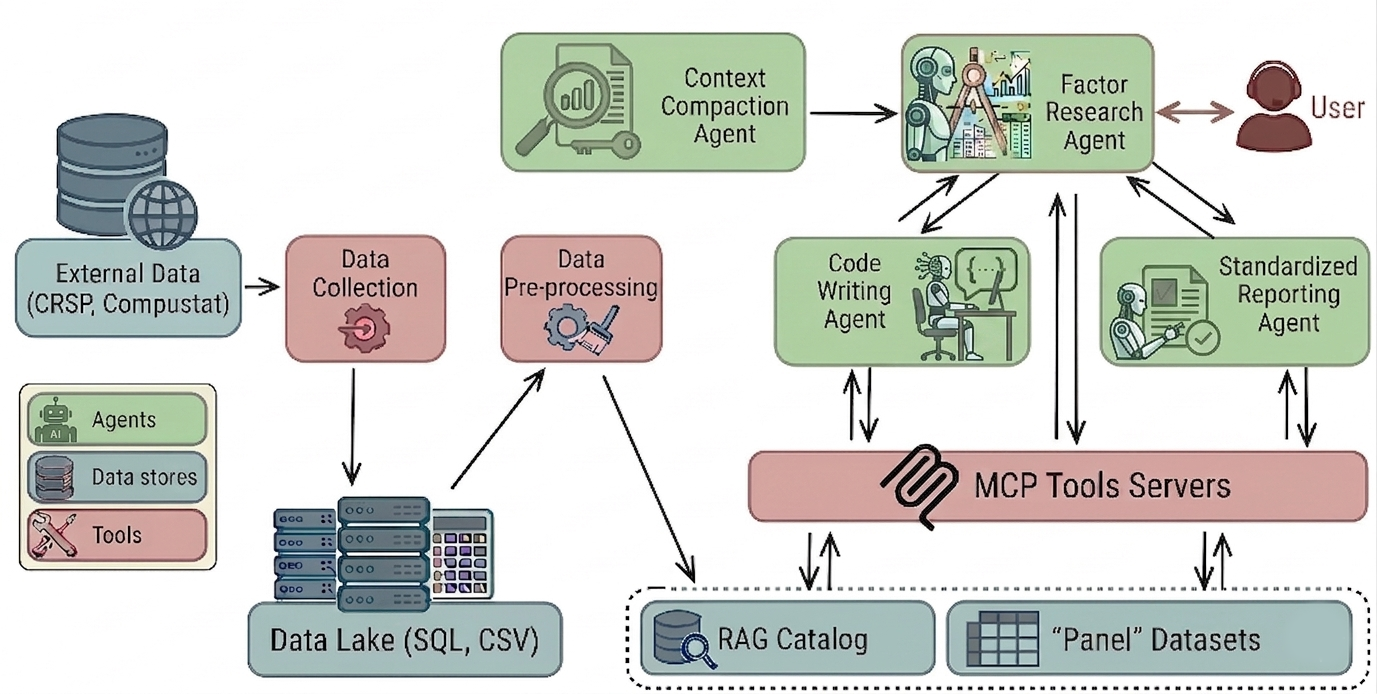}
  \caption{Agentic framework for empirical research in quantitative finance}
  \label{fig:architecture}
\end{figure*}

QRAFTI uses a multi-agent architecture in which specialized agents, shown in Figure~\ref{fig:architecture}, work together to emulate the workflow of a quantitative research team (with approximate human-equivalent roles shown in parentheses):

\begin{itemize}
\item \textbf{Factor Research Agent} (``Research Analyst''):
interprets the user's request, decomposes the task, selects tools, and maintains the factor-construction pathway.

\item \textbf{Standardized Reporting Agent} (``Risk Manager''):
generates standardized diagnostics and the narrative research report.

\item \textbf{Code Writing Agent} (``Quant Developer''):
dynamically produces and executes Python code when a task exceeds the available toolset or requires custom analysis or visualization.
\end{itemize}

This role specialization is motivated by multi-agent systems research and by evidence that specialized components can reduce error rates in complex tasks \citep{Sapkota2025}.

\section{Data collection and pre-processing}

\subsection{Data sources}
High-quality factor research commonly relies on CRSP (``Center for Research in Security Prices'') for stock returns and market equity, and on S\&P Compustat for accounting fundamentals, which are the same databases typically used in the studies we aim to replicate. Many academic and corporate institutions already subscribe to these databases, allowing researchers to access them, often through WRDS\footnote{Wharton Research Data Services (WRDS) was used in preparing this report. This service and the data available thereon constitute valuable intellectual property and trade secrets of WRDS and/or its third-party suppliers.}. Linking across data vendors requires stable identifiers (e.g., PERMNO), together with careful calendar alignment for other identifiers (e.g., GVKEY and CUSIP). QRAFTI's preprocessing pipeline applies these transformations so that downstream research tools operate on consistent panels of raw data.

\subsection{Data pre-processing}
The raw CRSP Stocks Monthly and Compustat Annual files are preprocessed into persisted \texttt{Panel} datasets, enabling QRAFTI to use the Panel API and MCP tools for downstream factor-research operations. Several issues specific to financial data commonly arise in this setting, such as survivorship bias, identifier changes, and corporate actions such as distributions and delistings. CRSP and Compustat already address many of these concerns. During preprocessing into panel datasets, all data items were aligned in time, and anomalous values---for example, negative quantities in contexts where only non-negative values are economically meaningful---were removed. Preprocessing also incorporated several conventions from the literature, including adjustments for delisting returns, share-class aggregation, and investment-universe screens \citep{shumway1997, bessembinder2026, fama1993common}.

A catalog lookup tool, using Retrieval-Augmented Generation (RAG), maps natural-language references to stock characteristics already collected and preprocessed as panel datasets, allowing users or agents to select the most relevant panel data items by identifier or description.

\section{Panel data class and API}
Many empirical asset-pricing computations can be expressed as operations on a $T \times N$ panel of observations indexed by time and asset. QRAFTI encapsulates this structure in a \texttt{Panel} class, allowing operations to be specified and analyzed at a higher level---for example, ``apply a cross-sectional transform each month'' or ``estimate a rolling trend statistic for every asset.'' By automating common panel-data operations such as winsorization, quantile binning, rolling transformations, and portfolio aggregation, the framework aims to reduce implementation variance across studies and simplify tool design.

\section{Tool calling}
Tools expose empirical-finance operators as documented callable functions that LLM agents can select and invoke while carrying out research tasks.

\subsection{Tool design}
A central design principle in QRAFTI is that a large share of factor research can be expressed through a relatively small set of reliable operators:
\begin{itemize}
  \item \textbf{Primitive operations:} binary and unary mathematical operators.
  \item \textbf{Cross-sectional transforms:} winsorization, standardization, quantile binning, and masking or universe restriction.
  \item \textbf{Time-series transforms:} rolling aggregations, lagging, and resampling at different frequencies.
  \item \textbf{Portfolio construction:} mapping characteristics to portfolio weights and portfolio weights to realized returns.
  \item \textbf{Reporting:} standardized evaluation templates aligned with replication protocols.
\end{itemize}

QRAFTI exposes these capabilities as callable tools through the Model Context Protocol (MCP), a standardized client-server architecture that helps reduce the overhead of linking heterogeneous systems. This design is particularly useful in empirical finance, where research workflows often rely on multiple external data sources and a consistent set of recurring empirical operations.

\subsection{Dynamic code generation and execution}
When built-in tools are insufficient, the Coding Agent writes and executes custom Python code drawing on concise documentation for the \texttt{Panel} API. In quantitative-finance agent frameworks, dynamic code generation expands the range of solvable tasks, but also introduces failure modes such as subtle bugs and incomplete workflows \citep{kundu2025multiaagentqf}.
To mitigate these risks, QRAFTI prefers tool calls whenever possible, while recording executed code and outputs for auditability.

\section{Planning and reflection}
Reflection-style planning employs a pattern in which a model does not simply generate a plan and execute it immediately. Instead, it first drafts a plan, critiques that plan, and then revises it before acting \citep{liu2024selfreflection, madaan2023selfrefine}. In QRAFTI, a self-reflection prompt operationalizes these ideas by requiring the Research Agent to pass through three distinct stages:

\begin{itemize}
\item \textbf{Phase 1 (Drafting):}
The agent proposes a sequential set of steps to answer the user's query---for example, a natural-language request describing how to construct a financial factor---using only the validated toolset and panel-data structures.

\item \textbf{Phase 2 (Critique):}
The agent audits the proposed plan to determine whether each step is technically implementable and whether the overall logic satisfies the research objective without introducing missing steps, invalid assumptions, or unnecessary inefficiencies.

\item \textbf{Phase 3 (Refinement):}
The agent produces a revised and corrected plan based on the preceding critique.
\end{itemize}

The key idea is to make self-critique an explicit intermediate step rather than concealing it within a single reasoning pass, reducing the risk that the Research Agent will prematurely commit to flawed logic.

\section{Context compaction}
Empirical research workflows often involve extended multi-step turns. LLMs may fail to recover relevant information embedded in long inputs \citep{liu-etal-2024-lost} or become ``lost'' during multi-turn conversations \citep{laban2025}. When the conversation history becomes too dense, a context-compaction step can be triggered to distill the most important information into a structured summary. This process uses a specialized LLM call to transform raw chat logs into a condensed representation that tracks each computed panel identifier, including its data inputs, transformations, and dependencies.

\section{Implementation}

The system uses the Pydantic-AI framework \citep{pydanticai2026}, which provides a type-safe environment ensuring that model outputs and tool parameters conform to strict schemas. Its integration with OpenTelemetry and Pydantic Logfire additionally enables detailed performance monitoring, cost tracking, and real-time tracing of agent trajectories. QRAFTI provides a user interface (UI) with interactive dashboards implemented in Streamlit. Computation graphs provide a visual representation of task dependencies and information flow across tool calls, where each constructed factor is represented as a directed acyclic graph from raw inputs to final outputs. It also includes a command-line interface (CLI) that allows users to interact with agents directly, without going through the UI. This allows the system to be integrated into scripts and automated workflows, enabling researchers to programmatically manage repeated experiments and collect results in a more systematic way.

We used Python 3.13 for development and set the temperature of Gemini 2.5-Pro to 0. The appendix provides visual examples of the Streamlit-based UI in operation.

\chapter[Evaluation]{Evaluation}

\section{Agentic evaluation}
When evaluating an agentic research system, the relevant object of study is an end-to-end workflow that includes planning, tool use, and reporting. Autonomous-agent frameworks have often been shown to fail because errors compound across multiple steps, underscoring the importance of evaluating such systems at the workflow level rather than solely in terms of single-turn accuracy \citep{yang2023autogpt}.

Inspection of QRAFTI's trace records and computation graphs helps assess whether the system correctly translates a natural-language query into a complete empirical specification, including data selection, lag structure, and factor operations. Consistency in the counts of rows and months with non-null values in intermediate and final data artifacts, relative to the reference, provides an additional audit indicator that the workflow is producing the intended shape, coverage, and temporal footprint.

Quantitatively, we assess QRAFTI by how closely the intermediate and final artifacts it generates---such as derived characteristics, quantile ranks, portfolio weights, and factor returns---match reference implementations.
\textit{Cosine similarity} treats series values as vectors and measures the size of the angle between them. Scores range from -1 to 1, with higher values indicating greater similarity.

\section{Experiments}

Our experiments employed QRAFTI agents to replicate empirical research designs from textual descriptions using tool calls and to write and execute custom code when built-in tools were insufficient.

\subsection{Replication from textual descriptions}
A canonical application is the replication of spread factors such as value (high book-to-market minus low book-to-market) or price momentum (past winners minus past losers). Given a natural-language description of the research specification including lag structure, breakpoints, quantile or standardization procedures, and weighting conventions, QRAFTI's Factor Research Agent generates an explicit tool-execution plan. Through reflection prompting, these plans may also be verified and corrected before execution.

\begin{itemize}
\item \textbf{HML-style workflow with multiple turns:}
replicates the HML value factor following the workflow described in \citet{fama1993common} and \citet{Hasler2021Value} with multi-turn queries shown in Figure~\ref{fig:hml}.

\begin{figure}[t]
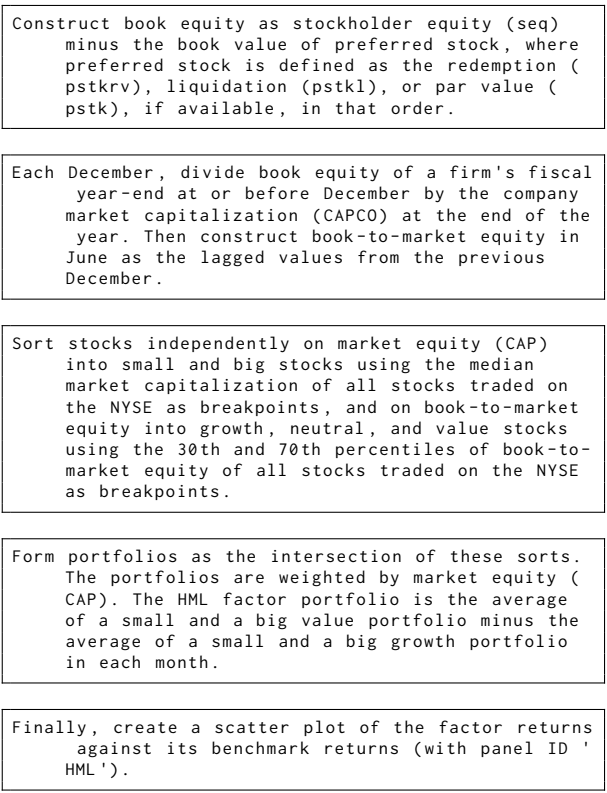

\begin{lstlisting}[language=plain]
Construct book equity as stockholder equity (seq) minus the book value of preferred stock, where preferred stock is defined as the redemption (pstkrv), liquidation (pstkl), or par value (pstk), if available, in that order.
\end{lstlisting}

\begin{lstlisting}[language=plain]
Each December, divide book equity of a firm's fiscal year-end at or before December by the company market capitalization (CAPCO) at the end of the year. Then construct book-to-market equity in June as the lagged values from the previous December.
\end{lstlisting}

\begin{lstlisting}[language=plain]
Sort stocks independently on market equity (CAP) into small and big stocks using the median market capitalization of all stocks traded on the NYSE as breakpoints, and on book-to-market equity into growth, neutral, and value stocks using the 30th and 70th percentiles of book-to-market equity of all stocks traded on the NYSE as breakpoints.
\end{lstlisting}

\begin{lstlisting}[language=plain]
Form portfolios as the intersection of these sorts. The portfolios are weighted by market equity (CAP). The HML factor portfolio is the average of a small and a big value portfolio minus the average of a small and a big growth portfolio in each month.
\end{lstlisting}

\begin{lstlisting}[language=plain]
Finally, create a scatter plot of the factor returns against its benchmark returns (with panel ID 'HML').
\end{lstlisting}
\caption {Fama-French HML-workflow user queries}
\label{fig:hml}
\end{figure}

\item \textbf{JKP-style price momentum workflow with reflection-based planning:}
replicates the price-momentum factor using the JKP-style workflow described in \citet{JensenKellyPedersen2023} with a single reflection-style query in Figure~\ref{fig:jkp}.
\end{itemize}

\begin{figure}[t]
  \begin{lstlisting}[language=plain]
Please use these three phases of the self-reflection prompt technique to perform the query below.
Phase 1: Consider the entire query, and suggest a sequential order of steps to perform the query.
Phase 2: To reflect and self-critique, check that each step is implementable with available tools and that the steps can efficiently satisfy the objective of the query; you may query the user to provide any needed definitions.
Phase 3: Provide the corrected plan, but do not execute the steps yet.

Query:
Define price momentum characteristic as stocks' past 12 months returns skipping one month.

Sort stocks into characteristic terciles (top/middle/bottom third) with breakpoints based on non-micro stocks, where micro stocks are all stocks whose market capitalization is below the NYSE 20th percentile.

For each tercile, compute its "capped value" weighted portfolio, meaning that we weight stocks by their market equity winsorized at the NYSE 80th percentile.

The factor returns are then defined as the top-tercile portfolio return minus the bottom-tercile portfolio return.

Create a scatter plot of the factor returns against its benchmark returns (Panel ID 'ret_12_1_ret_vw_cap').
  \end{lstlisting}
  \caption {Price momentum JKP-workflow user query}
 \label{fig:jkp}
\end{figure}

\subsection{Dynamic code generation and execution}
Not all workflows can be expressed through a fixed toolset. When the agent encounters a missing capability---for example, computing an exponentially weighted stock-volatility characteristic---QRAFTI delegates the task to a code-writing and execution agent. This agent writes Python code against the \texttt{Panel} API, executes it, and returns both the code and the resulting artifact (that is, the identifier of a newly created \texttt{Panel} dataset), which can subsequently be used in later tool calls. Reviewers may also inspect the generated code and test it deterministically.

\chapter[Results and Discussion]{Results and Discussion}

\section{Results of experiments}

\begin{figure}[t]
  \includegraphics[width=0.48\linewidth]{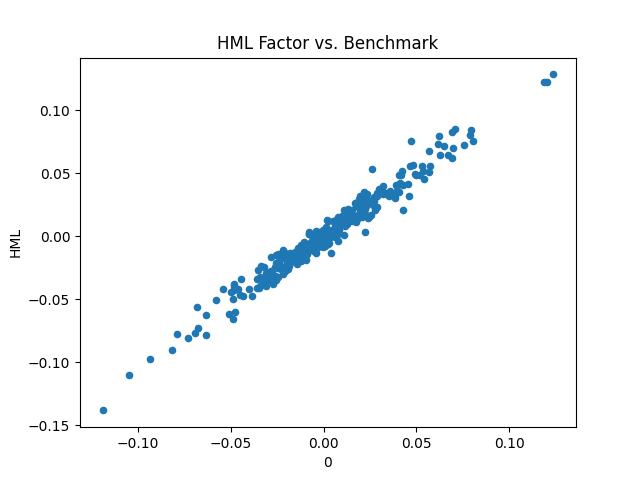} \hfill
  \includegraphics[width=0.48\linewidth]{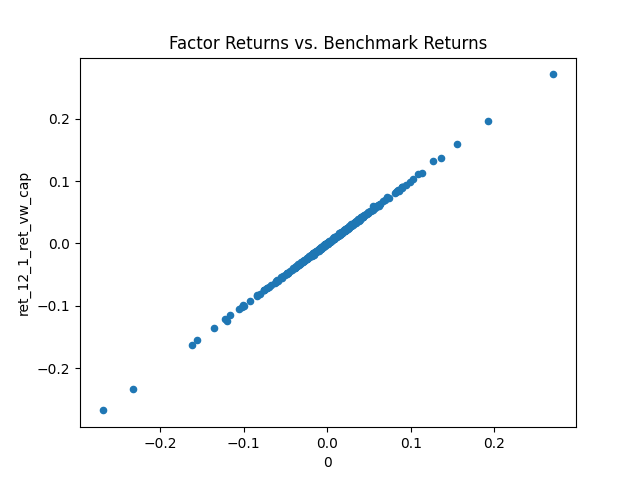}
  \caption {Scatter plots of constructed panels against reference benchmark factor returns}
  \label{fig:scatter}
\end{figure}

The final output artifacts from the two replication experiments were compared with benchmark factor-return series (named \texttt{HML} and \texttt{ret\_12\_1\_vw\_cap} respectively) obtained from the websites of Ken French\footnote{\url{https://mba.tuck.dartmouth.edu/pages/faculty/ken.french/data_library.html}} and \citet{JensenKellyPedersen2023}\footnote{\url{https://jkpfactors.com/}}. The scatter plots in Figure~\ref{fig:scatter} indicate strong agreement between the constructed panels and the reference benchmark returns, with observations lying close to the 45-degree line.

In the dynamic coding experiment, the system generated the custom code shown in Figure~\ref{fig:coding}, with
output that also matched the reference implementation.\footnote{Full conversation histories and outputs of the experiments are available in the Internet Appendix.}

\begin{figure}[t]
  \begin{lstlisting}[language=python]
from qrafti import Panel, DATES
import pandas as pd
import json

def ewma_helper(df: pd.DataFrame) -> pd.Series:
    """Computes the exponentially-weighted moving average for a series."""
    # Persistence factor = 1 - alpha
    # alpha = 1 - 0.94 = 0.06
    return df.iloc[:, 0].ewm(alpha=0.06, min_periods=12, adjust=False).mean()

# Load the input panel
input_panel = Panel().load("_164", **DATES)

# Apply the EWMA calculation for each stock using the trend method
result_panel = input_panel.trend(ewma_helper).save()

# Print the resulting panel's payload as a JSON string
print(json.dumps(result_panel.as_payload()))
  \end{lstlisting}
  \caption {Listing of custom code generated}
  \label{fig:coding}
\end{figure}

\section{Performance by task complexity and configuration}
We also conduct ablation studies to examine performance under two restrictions:
using only dynamic code generation (i.e., without access to the broader toolset), and
removing reflection-based planning prompts for multi-step tasks.

We define \textbf{1-step tasks} as the first instruction in each multi-step workflow query, \textbf{2-step tasks} as the first two instructions, and \textbf{multi-step tasks} as the full workflow instructions from the two replication examples.

In these experiments, each agent is asked to answer every query five times, which allows us to evaluate both accuracy and consistency across repeated trials. For each task, we report \textit{Sim@k} for \(k \in \{1,2,5\}\), defined as the expected maximum cosine similarity between the reference panel and a set of \(k\) generated outputs sampled from the \(n\) attempts. This metric is similar in spirit to \textit{Pass@k} \citep{kulal2019}, but replaces exact correctness with cosine similarity to a benchmark reference panel. Here, we relax the requirement of an exact match because outputs may differ slightly from benchmark reference panels even when they are substantively correct, owing to minor implementation choices and the large underlying and output datasets. Specifically, for task $i$ with output $y_{i,j}$ on the $j$-th of $n$ attempts:

\[
\mathrm{sim@}k_i
=
\frac{1}{\binom{n}{k}}
\sum_{S \in \mathcal{S}_{n,k}}
\max_{j \in S}
\cos\!\left(y_{i,j},\, y_i^{\mathrm{ref}}\right),
\]

where \(\mathcal{S}_{n,k}\) denotes the set of all size-\(k\) subsets of \(\{1,\dots,n\}\), and \(\cos(\cdot,\cdot)\) denotes the cosine similarity.  The final metric is the average across $N$ tasks:

$$\mathrm{Sim@k} = \frac{1}{N} \sum_{i=1}^{N} \mathrm{sim@k}_i$$

Hence, analogous to Pass@k, Sim@k measures the expected quality of the best
result from \(k\) randomly selected attempts, where quality is defined by cosine
similarity with the reference panel rather than exact correctness.

\begin{table*}[ht]
  \centering
  \small
\begin{tabular}{llccc}
\hline
Task Complexity & Configuration & \textit{Sim@1} & \textit{Sim@2}  & \textit{Sim@5}
\\
\hline
1-step    & All tools                  & 0.9675 & 0.9881 & 1.0000 \\
          & Coding-tool only           & 0.7213 & 0.9330 & 1.0000 \\ \hline
2-step    & All tools                  & 1.0000 & 1.0000 & 1.0000 \\
          & Coding-tool only          & 0.5910 & 0.8281 & 0.9826 \\ \hline
Multi-step & All tools                  & 0.7435 & 0.9263 & 0.9760 \\
  & All tools + reflection-planning & 0.9858 & 0.9908 & 0.9927 \\ \hline
\end{tabular}
\caption{Average performance by task complexity and configuration}
\label{tab:results}
\end{table*}

Table~\ref{tab:results} shows that access to the full toolset improves performance as task complexity increases, while reflection-based planning is important on multi-step workflows.
For 1-step and 2-step tasks, the all-tools configuration achieves near-perfect performance from the first attempt, whereas the coding-tool-only configuration performs worse, especially on 2-step tasks, where Sim@1 and Sim@2 are only 0.59 and 0.83. In the full multi-step setting, using all tools without reflection-planning yields only moderate one-shot performance, though repeated attempts improve results substantially, while adding reflection-based planning raises performance from the first attempt.

\section{Autonomous generation and testing of new factors}
Beyond replication, QRAFTI can also serve as an ``idea-to-evidence'' engine. It can generate candidate factor ideas from a high-level prompt, implement them rapidly, and evaluate them using the same standardized diagnostic framework applied to established factors. Given a broad hypothesis, the agent proposes one or more candidate implementations expressed as explicit formulas using available variables. It then translates these formulas into tool calls or code and produces a complete empirical evaluation by delegating to the standardized research-report agent. As an example, we provide QRAFTI with an open-ended query instructing it to suggest and test a factor reflecting ``the investment philosophy of the renowned investor Warren Buffett.'' The system responded by constructing a standardized combination of quality and value. The quality side blends four signals: profitability via return on equity, gross profitability (gross profits relative to assets), low stock-return volatility over a 36-month window, and low leverage measured by debt-to-assets; lower volatility and leverage are inverted so that higher scores always indicate better quality. All accounting variables are lagged six months, then aggregated first into a Quality Score and a Value Score, and finally into a single ``Buffett-style'' quality-at-a-reasonable-price factor. These factor construction steps are represented as a computation graph through the UI. The Appendix also shows that a tercile spread implementation strategy had a negative Sharpe ratio, but a positive alpha after controlling for its inverse sensitivity to the market.

Evaluating proposed factors as genuinely \emph{investable} signals remains an open challenge in empirical asset pricing \citep{feng2020taming}. Pure out-of-sample validation is difficult in real time. A truly clean out-of-sample test often requires waiting years for new data \citep{french2022five, baz2024alpha}, while post-publication performance may also decay because of both data-snooping and market adaptation \citep{lo1990data,mclean2016}. Furthermore, results are often sensitive to implementation details and assumptions about transaction costs. These concerns motivated \citet{novy-marx2023assaying} to propose a diagnostic protocol that reduces researcher degrees of freedom and makes the evaluation of new signals more transparent, replicable, and economically interpretable. We adapt their framework to design a report that provides a more structured assessment of whether a candidate signal is likely to reflect a real and usable source of return, rather than a fragile in-sample artifact.

\subsection{Standardized research reporting}
QRAFTI's Report Agent produces these standardized diagnostics along with a narrative summary, as shown in the Appendix. These outputs should be read together. A signal is more convincing when it performs well across multiple diagnostics, while strong results in one table should be viewed cautiously if other sections show weak coverage, microcap concentration, high turnover costs, or sensitivity to small implementation choices.

\begin{itemize}
    \item \textbf{Coverage by Period.}
    This section summarizes subperiod coverage using metrics such as the fraction of securities represented and the share of total market capitalization covered. A signal that works only in narrow windows or on a small subset of stocks is less credible as a general return predictor.

    \item \textbf{Spread Portfolio Summary Statistics.}
    This section reports descriptive statistics for the long-short return spread formed on the candidate characteristic. These statistics give an initial sense of economic magnitude and return distribution, but a high average return or Sharpe ratio is not enough if it vanishes after risk adjustment or depends mainly on small stocks.

    \item \textbf{Alpha, Coefficients, and t-Statistics by Model.}
    This section reports intercepts and factor loadings from time-series regressions under alternative benchmark models. These results help assess whether the spread portfolio appears to earn abnormal returns after controlling for common risks. Given the multiple-testing concerns emphasized by \citet{harvey2016crosssection}, a new signal is more convincing when its risk-adjusted alpha remains comfortably above a t-statistic of about 3.0, not just 1.96.

    \item \textbf{Alpha and t-Statistics by Model and Size Quantile.}
    This section breaks the evidence down by firm size. These diagnostics help show whether performance, measured by average returns and alphas under alternative benchmark models, is concentrated in smaller, less liquid stocks. Many anomalies are strongest where arbitrage is limited and trading costs are highest \citep{lam2011, lam2020}. A signal is more credible as an investable strategy when its performance is not limited to the smallest firms.
\end{itemize}

\chapter{Conclusion}
The integration of agentic AI could alter how quantitative research workflows are executed. The researcher’s role may shift away from manual data manipulation and coding toward the design and management of agentic research assistants. In this setting, AI could take on repetitive tasks in factor construction, portfolio implementation and preliminary analysis, allowing human researchers to focus more on research design, validation, and economic interpretation.

QRAFTI provides a practical foundation for moving from chat-style LLMs toward tool-grounded, multi-agent systems for empirical finance. By combining a panel-data abstraction, MCP-exposed research tools, computation-graph traceability, standardized reporting, and dashboard-based monitoring, it demonstrates how LLM-based agents can support quantitative research workflows while emphasizing reproducibility and explainability. More extensive evaluation across a wider range of tasks, datasets, and configurations would provide a better assessment of robustness and generalizability.

\chapter*{Disclaimer}
The views expressed here are those of the authors alone and not of Graphen, Inc.


\bibliography{references}

\appendix

\chapter{Appendix}

\section{User interface demonstration}
\begin{figure*}[t]
  \includegraphics[width=0.42\textwidth]{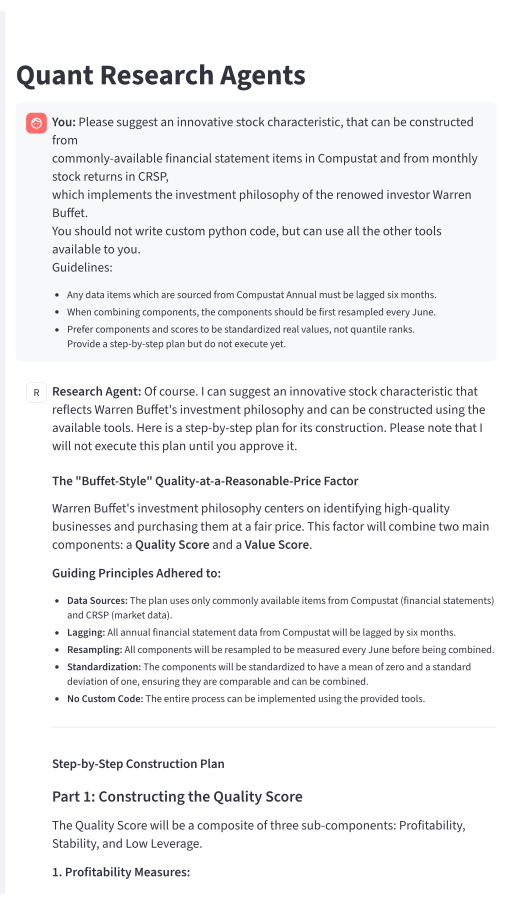} \hfill
  \includegraphics[width=0.44\textwidth]{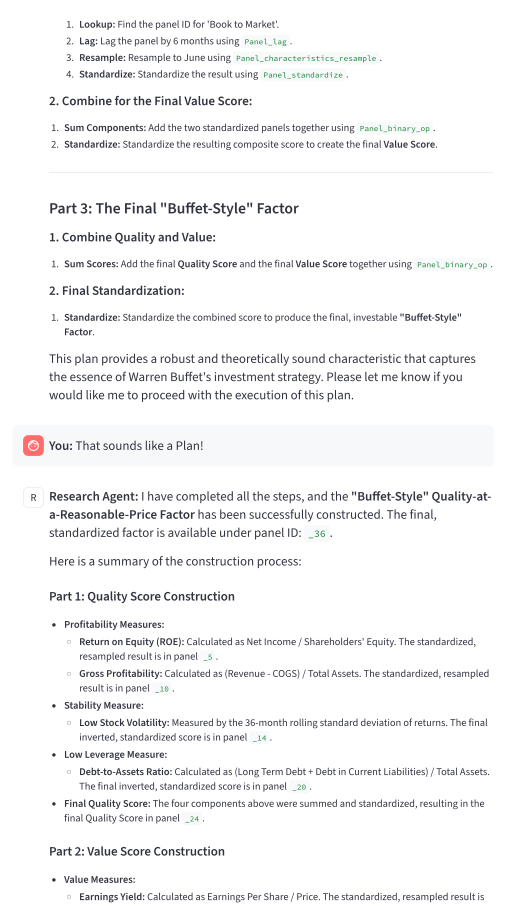}
  \includegraphics[width=0.42\textwidth]{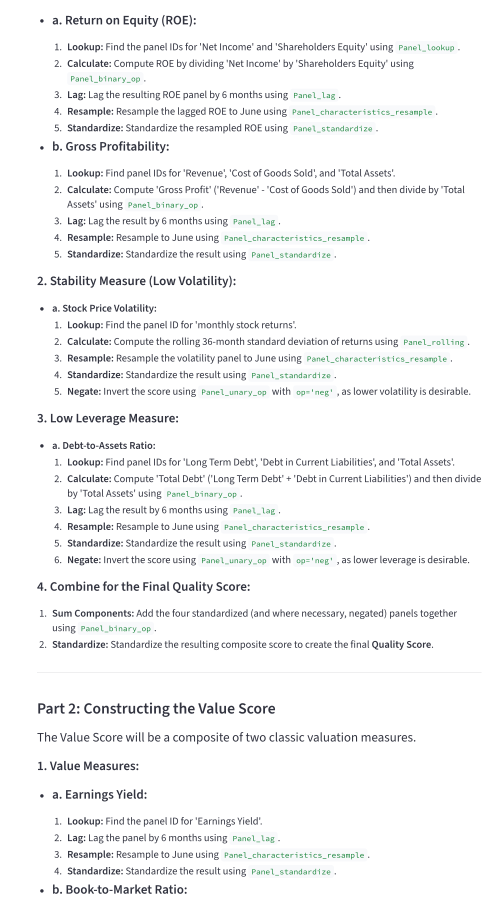} \hfill
  \includegraphics[width=0.44\textwidth]{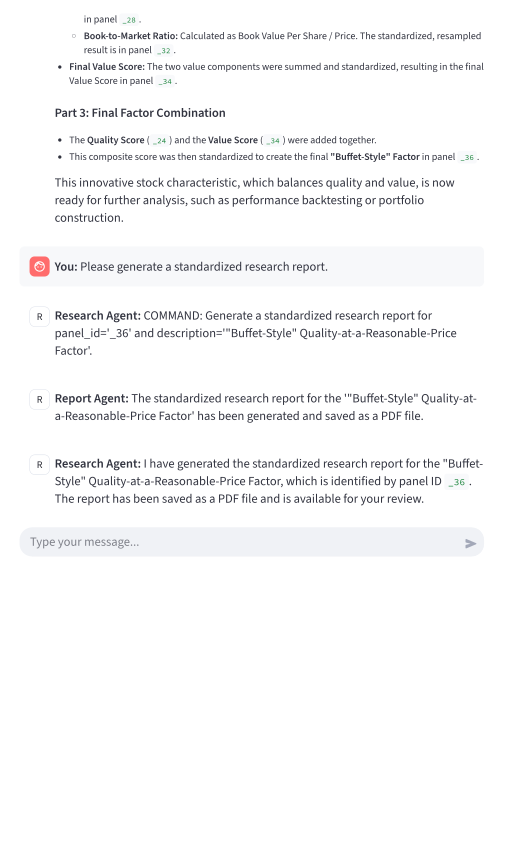}
  \caption {UI Demo: Conversation history}
  \label{fig:UI}
\end{figure*}

\begin{figure*}[t]
  \includegraphics[width=0.46\textwidth]{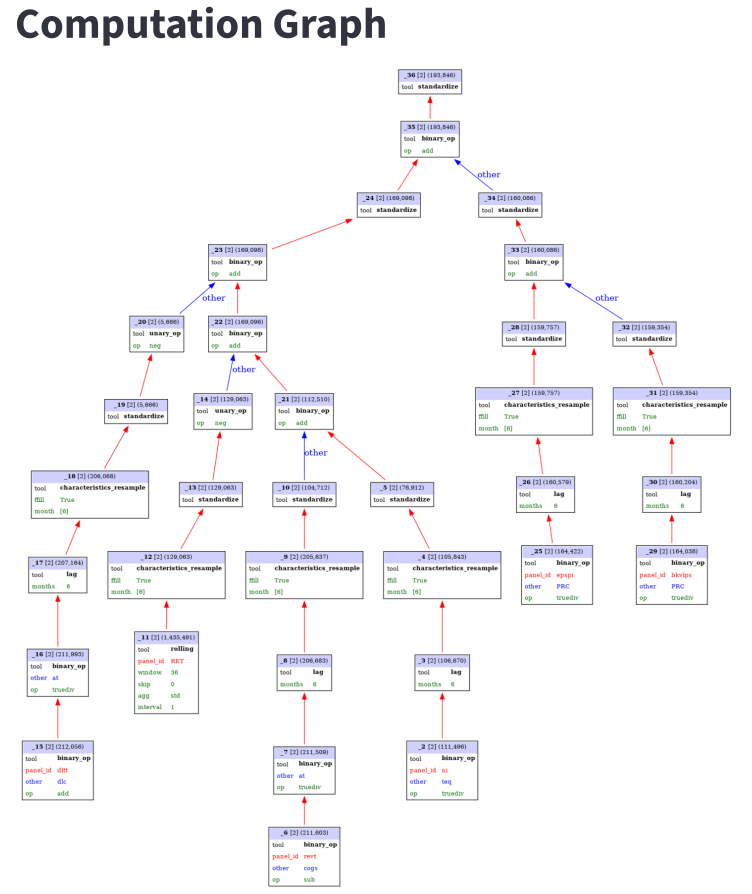} \hfill
  \includegraphics[width=0.46\textwidth]{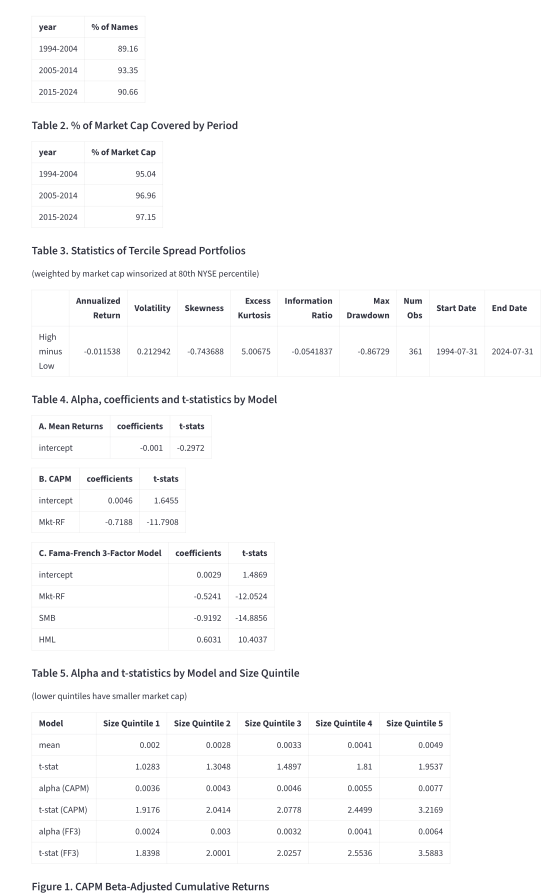}
  \includegraphics[width=0.46\textwidth]{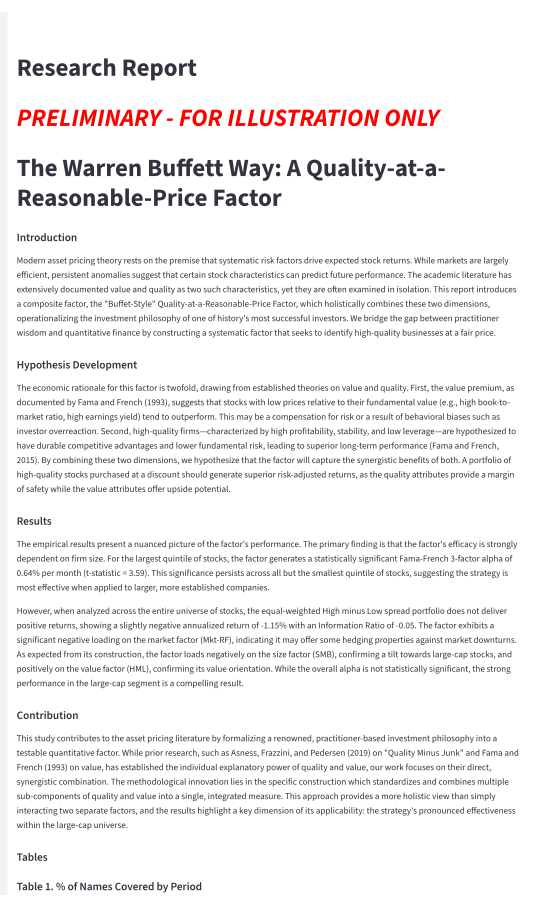} \hfill
  \includegraphics[width=0.46\textwidth]{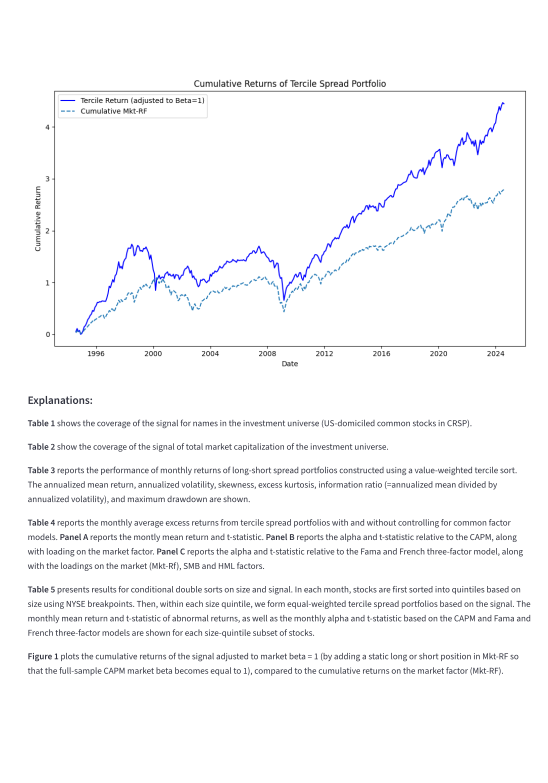}
  \caption {UI Demo: Computation graph and standardized research report with narrative analysis}
  \label{fig:display}
\end{figure*}

Figures \ref{fig:UI} and \ref{fig:display} display user interaction with QRAFTI, showing the
conversation histories and outputs generated, including a computation graph and
standardized diagnostics report.

In this session, the user began with an open-ended prompt asking the
system to suggest and test a factor that reflects the ``investment
philosophy of the renowned investor Warren Buffett.'' The system
responded with a description of a composite factor it had hypothesized,
the data items it would retrieve, and the steps required to compute the
intermediate and final panels. Upon the user's request, the system
presented a standardized diagnostic report together with narrative
analysis, as well as a computation graph that visually represents
dependencies and information flow across tool calls. Each constructed
factor is shown as a directed acyclic graph, tracing the process from
raw inputs to final outputs.

\end{document}